\documentclass[twocolumn,superscriptaddress,a4paper,prd]{revtex4-1}
\usepackage{amssymb,amsmath,latexsym,mathrsfs,color}
\usepackage{graphicx}

\newcommand{\beq}{\begin{equation}}
\newcommand{\eq}{\end{equation}}
\newcommand{\bear}{\begin{eqnarray}}
\newcommand{\ear}{\end{eqnarray}}

\begin{document}

\title{Scaling properties of multitension domain wall networks}

\author{M. F. Oliveira}
\email[]{migueloliveira@fc.up.pt}
\affiliation{Centro de Astrof\'{\i}sica, Universidade do Porto, Rua das Estrelas, 4150-762 Porto, Portugal}
\affiliation{Faculdade de Ci\^encias, Universidade do Porto, Rua do Campo Alegre 687, 4169-007 Porto, Portugal}
\author{C. J. A. P. Martins}
\email[]{Carlos.Martins@astro.up.pt}
\affiliation{Centro de Astrof\'{\i}sica, Universidade do Porto, Rua das Estrelas, 4150-762 Porto, Portugal}
\affiliation{Instituto de Astrof\'{\i}sica e Ci\^encias do Espa\c co, CAUP, Rua das Estrelas, 4150-762 Porto, Portugal}

\begin{abstract}
We study the asymptotic scaling properties of domain wall networks with three different tensions in various cosmological epochs. We discuss the conditions under which a scale-invariant evolution of the network (which is well established for simpler walls) still applies, and also consider the limiting case where defects are locally planar and the curvature is concentrated in the junctions. We present detailed quantitative predictions for scaling densities in various contexts, which should be testable by means of future high-resolution numerical simulations.
\end{abstract}
\date{26 January 2015}
\pacs{}

\maketitle
%%%%%%%%%%%%%%%%%%%%%%%%%%%%%%%%%%%%%%%%%%%%%%%%%%%%%%%%%%%%%%%%%%%%%%%%%%%%%%%%%%%%%%%%%%%%%%%%%%%%%
\section{Introduction}
\label{intr}

Phase transitions in the early universe have a number of inevitable consequences, including the formation of topological defects via the Kibble mechanism \cite{Kibble,Book}. The more interesting among these are cosmic strings, for which recent developments in fundamental theory have led to a resurgence of interest. Specifically, it was noticed that the production of cosmic (super) strings is much more generic than previously thought. Recent reviews of this topic \cite{CopelandKibble,CopelandPogosian} provide additional details of these developments. The potential observability of these strings can open a window into physics at the highest energy scales.

These objects are much more complex than ordinary field theory strings. Studying their evolution is a challenging problem involving physics from different energy scales, which is often not fully understood. Quantitative analytic models, even for the simplest strings \cite{ms1b,extend}, necessarily rely on averaging over some network configurations, which require the introduction of phenomenological parameters whose values can only be inferred by using numerical calibration \cite{Moore,Unified,Ring,Olum}. Various types of well-motivated string networks remain whose cosmological evolution is relatively unexplored, including cosmic strings carrying charges or currents \cite{charge} (which arise naturally in SUSY models) and models with junctions connecting string segments with different tensions, though there has been recent progress in the latter \cite{Urrestilla,Pourtsidou} and some relevant numerical studies have been carried out for domain walls \cite{Junctions}.

A more modest but useful approach is to start by studying simple toy models, where it may be easier to understand the various dynamical mechanisms at play, and then using this knowledge for tackling more complicated (but also more realistic) scenarios. Domain walls, being the simplest defect (in the sense that they can be described by a single scalar field) provide a particularly useful testbed, despite being tightly constrained by observations \cite{Zeldovich,Junctions}---though even this case has its challenges and requires large computing resources \cite{Leite1,Leite2}.

Here we study the asymptotic scaling properties of domain wall networks with multiple tensions, by generalizing an analytic model for the simplest walls. Our main interest is on the conditions under which a scale-invariant evolution of the network (whose presence for the simplest walls is well established, both on analytic and numerical grounds) still applies, but we will also consider a physically interesting low-curvature limit. The goal is to use this information to shed light on the analogous case for cosmic strings, in which case the issue of the generality of the scale-invariant asymptotic evolution is still not fully clear, although several analytic \cite{Tye,Tasos,Pourtsidou} and numerical efforts \cite{McGraw,Hindmarsh,Rajantie,Urrestilla} suggest that it applies in a wide range of circumstances.

\section{Analytic model}
\label{model}

An analytic model of a defect network starts with the microscopic equations of motion (the Nambu-Goto equations, in the case of strings) and through a suitable averaging process arrives at 'thermodynamic' evolution equations. These must include phenomenological terms to account for defect interactions and energy losses, and for their calibration one must resort to numerical simulations. For $U(1)$ cosmic strings, one obtains the velocity-dependent one-scale (VOS) model \cite{ms1b,extend}, which has been thoroughly tested against simulations. Analogous models have also been obtained for monopoles \cite{Monopoles} and for domain walls \cite{MY2}. In what follows, we provide a concise derivation of the analytic model for domain walls, summarizing the results of \cite{MY2}.

Let us start by considering a flat Friedmann-Robertson-Walker (FRW) universe containing a network of non-interacting infinite planar domain walls, oriented along some fixed direction and all with the same value of the velocity $v$. The momentum per unit comoving area of the walls is proportional to $a^{-1}$ (where $a$ is the cosmological scale factor) so that we have $v \gamma \propto a^{-3}$. Differentiating and noting that the Hubble parameter is $H=(1/a)(da/dt)$, we have
\begin{equation}
\frac{dv}{dt}+3H(1-v^2)v=0\,.
\label{vevoldw}
\end{equation}
In the absence of interactions between walls, the average number of domain walls in a fixed comoving volume should be conserved so that $\rho \propto \gamma a^{-1}$ or equivalently
\begin{equation}
\frac{d \rho}{dt}+H(1+3v^2)\rho=0\,,
\label{rhoevoldw}
\end{equation}
where $\rho$ is the average energy density in domain walls and we have again used Eqn. (\ref{vevoldw}) to obtain (\ref{rhoevoldw}).

Now, instead of assuming the domain walls to be infinite and planar we may define a characteristic length scale,
\begin{equation}
L={\frac{\sigma}{\rho}}\,,
\end{equation}
which is directly related to the average distance between adjacent walls, with $\sigma$ being the domain wall mass per unit area. In cosmologically relaistic domain wall networks this characteristic length scale $L$ will be approximately equal to the curvature scale of the walls, and the walls will have a non-zero probability of crossing and interacting with each other. Moreover, the value of the velocity will vary along the walls; nevertheless, in the absence of interactions Eqn. (\ref{rhoevoldw}) would remain valid, with $v$ being taken as the RMS velocity of the strings.

We therefore need to add further terms to Eqns. (\ref{vevoldw}) and (\ref{rhoevoldw}) in the context of realistic models. Let us consider the latter equation first. The probability of a wall element of characteristic size $L$ encountering another segment of the same size within a time $dt$ is proportional to $vdt/L$ \cite{Book}. We thus expect an energy loss rate
\begin{equation}
\left(\frac{d \rho}{dt}\right)_{\rm loss}= -c\frac{v}{L}\rho\,,
\label{looplosses}
\end{equation}
where we have introduced a dimensionless proportionality factor, $c$, which we may expect to be a constant. (This expectation is in agreement with numerical simulation results.) We may therefore add this term to the right hand side of (\ref{rhoevoldw}), which will account for the energy lost from the wall network due to the production of wall blobs---also called 'vacuum bags'. This mechanism is analogous to loop production in cosmic string networks, although the vacuum bags will typically dacay much faster than the string loops \cite{Zeldovich}. The evolution equation now becomes
\begin{equation}
\frac{d \rho}{dt}+H(1+3v^2)\rho=- c\frac{v}{L}\rho=-\frac{cv}{\sigma}\rho^2\,,
\label{rhoevol1}
\end{equation}
or, writing it in terms of the length scale $L$,
\begin{equation}
\frac{dL}{dt}=(1+3v^2)HL+cv\,  \label{rhoevol}
\end{equation}
This assumes that walls do intercommute when they interact, but the phenomenological parameter $c$ models the overall energy losses and can account for the limiting case where there are no losses (by setting $c\to0$).

On the other hand, we also assumed planar infinite walls. Realistic wall networks will of course be curved, and their curvature will be responsible for an acceleration term which also needs to be taken into account. The acceleration should be inversely proportional to the curvature radius, and in the context of a one-scale model the curvature radius is identified with $L$ itself. To account for the fact that this identification, although approximately valid, need not hold exactly, a further dimensionless curvature parameter $k$ (expected to be of order unity) is also introduced. The velocity equation is thus corrected to
\begin{equation}
\frac{dv}{dt}=(1-v^2)\left(\frac{k}{L}-3Hv\right)\,. \label{vevol}
\end{equation}
Note that physically the evolution equation for $L$ (or the density) is an energy conservation equation, while that for $v$ is analogous to Newton's second law.

The above derivation can be analogously done for strings, in which case on recovers the evolution equations that can be independently obtained startng from the microscopic equations of motion derived from the Goto-Nambu action \cite{ms1b,extend}. This, together with comarisons with numerical simulations, lends further support to the validity of these evolution equations. We emphasize that this is a {\it one-scale model}, meaning that one assumes that there is a single relevant lengthscale in the problem. In addition to the characteristic lengthscale $L$ defined above (which is essentially a parametrization of the energy density in the wall network) one can define a correlation length $\xi$ and a curvature radius $R$, for example. In the context of this  model, one is assuming that $L=\xi=R$. (In the case of cosmic strings, high-resolution Goto-Nambu numerical simulations have shown that this is a good approximation \cite{Moore}.)

The model therefore has two free parameters, $c$ and $k$: the former quantifies (fractional) energy losses by the network, while the latter quantifies the curvature-related forces acting on the walls. To a first approximation these are expected to be constant, and this has been confirmed in recent high-resolution numerical simulations \cite{Leite1,Leite2}, which found
\begin{equation}
c=0.34\pm0.16\,,\qquad k=0.98\pm0.07\,. \label{newkw}
\end{equation}
Effectively, these provide a numerical calibration for the analytic model.

Neglecting the effect of the wall energy density on the background (specifically, on the Friedmann equations), which is the usually the case when one carries out high-resolution field theory simulations of these networks, one can show that the attractor solution to the evolution equations (\ref{rhoevol},\ref{vevol}) is a linear scaling solution
\begin{equation}
L=\epsilon t\,, \qquad v=const\,. \label{defscaling}
\end{equation}
Assuming that the scale factor behaves as $a \propto t^\lambda$, the linear scaling parameters have the following detailed form:
\begin{equation}
\epsilon^2=\frac{k(k+c)}{3 \lambda (1-\lambda)}\,,\qquad v^2=\frac{1-\lambda}{3\lambda}\frac{k}{k+c}\,. \label{scaling2}
\end{equation}
In practice, when carrying out numerical simulations one looks for the best fit to the power laws $A/V\propto\rho_w\propto1/\xi_c\propto\eta^{\mu}$ and $\gamma v\propto\eta^{\nu}$, where $\xi_{c}$ is the comoving correlation length (the numerical counterpart of the physical lengthscale $L$), $A/V$ is the fraction of the simulation box containing walls, and $\gamma$ is the Lorentz factor. For a scale-invariant behavior, we should have $\mu=-1$ and $\nu=0$. 

\section{Multiple tensions}
\label{mult}

We now consider a model with domain walls of 3 different types, with tensions (ie, wall masses per unit area) $\sigma_1$, $\sigma_2$, $\sigma_3$. The purpose of this model is to describe the interaction between the different types of defects, in order to determine if this kind of networks can reach a scaling regime, as suggested in \cite{Tye,Tasos}. Additional motivation for working with three types of walls is also provided by \cite{Tasos}, who find that in the case of cosmic strings the three lightest components dominate the dynamics, with the heavier ones being suppressed.

In describing a network with 3 types of walls, one might start by using a set of evolution equations for $L$ (or $\rho$) and $v$ for each of the wall types. In fact, for many purposes one can simply use a single, averaged velocity for the entire network; we will comment further on this assumption shortly. More importantly, one must have further phenomenological terms describing the interaction---specifically, the energy transfer---between the three types of walls. This will also lead to the appearance of additional parameters in the model.

For simplicity, in this model we consider that from the 3 types of walls, types 1 and 2 behave as standard wall networks, with interactions between defects of the same type being modelled as in the standard case. But, when walls of type 1 collide with those of type 2 (or vice versa), they produce a segment of a new type of wall, type 3 (with tension $\sigma_3$). This behaviour requires the introduction of an additional term, modelling the interaction between walls of types 1 and 2.

We can now model this system by generalizing the one-scale model description. Specifically, using the results of the previous section (and in particulat the fact that $\rho_i=\sigma_i/L_i$), the evolution equations for the densities of the three types of walls and for the average network velocity become
\begin{equation}
\frac{d\rho_i}{dt}=-(1+3v^2)H\rho_i-\frac{cv}{\sigma_i}\rho_i^2-\frac{1}{2}\dot{\rho}_{1,2\rightarrow3}
\end{equation}
for $i=1,2$;
\begin{equation}
\frac{d\rho_3}{dt}=-(1+3v^2)H\rho_3-\frac{cv}{\sigma_3}\rho_3^2+\dot{\rho}_{1,2\rightarrow3}
\end{equation}
and
\begin{equation}
\frac{dv}{dt}=(1-v^2)\left(\frac{k \rho_3}{\sigma_3}-3Hv\right)\,.
\end{equation}

These are therefore the usual evolution equations for each type of wall (through written in terms of the density rather than correlation length, cf. the previous section), except that a common velocity has been assumed for all of them (more on this below) and a phenomenological interaction term between the different types was added, which in our case describes the production of wall segments of type-3 as a result of the collision of type-1 and type-2 walls. All that remains to be done is therefore to obtain the specific form of this term.

Note that above we have assumed that type-1 and type-2 networks have equal energy losses as the result of the interactions. This assumption is reasonable for networks with comparable tensions (which is likely to be the realistic case). Its validity is less clear in the case where one tension is much larger than the other. In that case, if both networks lose comparable lengths, then the lighter one will clearly lose less rest energy. On the other hand, the lighter network should be moving faster, implying greater losses from kinetic energy. The relative contributions of the two effects may well depend on both energetic and topological considarations, and therefore be model-dependent. Such a detailed analysis is beyond the scope of our work, but we do caution the reader that this equipartition assumption should be tested with numerical simulations.

We will need to introduce a new dimensionless phenomenological parameter, which we'll denote $d$: an efficiency parameter controlling the rate of energy transfer into type 3 walls, much in the same way the $c$ parameter does for loops (self interaction between walls of the same type) \cite{ms1b}. Thysically this term should be symmetric in $L_1$ and $L_2$ (or $\rho_1$ and $\rho_2$) and grow with both of these densitities, since higher densities of types 1 and 2 will lead to more interactions, as will a higher average velocity of the network. For the same reason, it should decrease as one increases the density of walls of type 3, $\rho_3$. For the case of strings, this energy loss term was shown in \cite{Tasos} to have the form
\begin{equation}
{\dot\rho}\propto E_3/(L_1^2L_2^2)\,
\end{equation}
with $E_3$ being the energy (of type-3 defects) produced in a volume $L_3^3$ per unit time. On dimensional grounds this behaviour should apply for walls, the only salient difference being that for the one-scale model for strings the correlation length, density and string tension $\mu$ are related by $\rho_{string}=\mu/L^2_{string}$ while for walls we have $\rho=\sigma/L$. Using the latter definition we can therefore write the energy transfer term as
\begin{equation}
\dot{\rho}_{1,2\rightarrow3}=d\, v\, \frac{\sigma_3^3}{\sigma_1^2\sigma_2^2} \frac{\rho_1^2\rho_2^2}{\rho_3^2}.
\end{equation}
The only caveat is that the phenomenological value for the parameter $d$ inferred from string simulations need not hold for walls, and in what follows we will therefore treat $d$ as a free parameter, though with the expectation that it be of order or slightly less than unity. We note, in passing, that the values inferred from simulations for the model parameters $c$ and $k$ are also different for strings and walls.

At this point it's useful to summarize the underlying assumptions in our model for multi-tension walls, and to compare them with those in existing models for multi-tension strings. Indeed, our assumptions do differ slightly from previous works for strings \cite{Tye,Tasos}. While \cite{Tye} assumes a single correlation length and velocity for the entire network (allowing for different densities for each string type), \cite{Tasos} assumes different correlation lengths and velocities for each type (and relates correlation lengths and densities through a Brownian assumption). Our assumptions are therefore intermediate between these two.

It must also be emphasized that the terms describing the interactions between the three wall types satisfy energy conservation. This assumption (not explicitly enforced by \cite{Tye}) is a conservative approach when it comes to scaling. For simplicity we only consider energy transfers from two types of walls to the third one, but we have checked that our general results would still hold if all possible energy transfers were allowed. In other words, the broad qualitative features of the solutions (whether or not there is scaling, for example) would remain unchanged, although the exact values of the scaling densities and velocities would of course depend on the numerical values of the parameters. Since the former are our main concern here, we have chosen to concentrate on the simpler case: discussing a more general one would mean added mathematical complexity but little additional physical insight.

\begin{figure}
\includegraphics{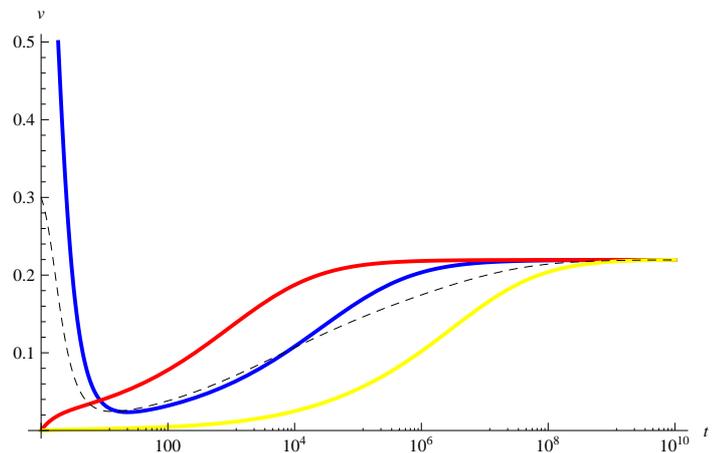}
\caption{\label{vwalls}Example of the evolution of the individual velocities of the walls of type 1 (blue), 2 (red), and 3 (yellow). The dashed line represents the average velocity of the network, showing it to be a good approximation.}
\end{figure}

The use of a single average velocity for the network is a good approximation for the velocity of any of its components. This is expected from previous work, but we have verified that using different velocities for each type of wall yields the same qualitative results as the ones we'll describe. A simple comparison of the two approaches is shown in Fig. \ref{vwalls} (with all other model parameters being the same in both cases). Asymptotically all three components reach a common velocity, although the timescale in which they do so is different for different wall types. For the purposes of the current work, which is concerned with asymptotic scaling properties, an average velocity for the network is an excellent approximation. (One would only need to go beyond this if calculating detailed astrophysical observables for comparison with experiments.)

Along these lines, note that assuming that the phenomenological energy loss parameter $c$ is the same for all three networks is a reasonable assumption. On the other hand, if we were to allow for multiple velocities there would be no reason for assuming equal curvature parameters $k$ for each string type (so allowing for multiple velocities would greatly increase the number of free parameters).

Having the evolution equations for the network, we can proceed to finding its scaling solutions, if they exist. We look for solutions of the type
\begin{equation}
L_1=\frac{\sigma_1}{\rho_1}=\epsilon_1 t^\alpha\,\quad L_2=\frac{\sigma_2}{\rho_2}=\epsilon_2 t^\beta\,\quad L_3=\frac{\sigma_3}{\rho_3}=\epsilon_3 t^\gamma
\end{equation}
\begin{equation}
v=v_o t^\delta\,.
\end{equation}
The result of this analysis is that two branches of solutions are allowed, with the parameter distinguishing them being the curvature parameter $k$. We now discuss them in turn.

The first type of solution corresponds to $k\neq0$, and is characterized by
\begin{equation}
\alpha=\beta=\gamma=1\,,\quad \delta=0\,,
\end{equation}
that is, all walls scale linearly with time, while the velocity goes to a constant value $v_0$, which can be determined based on the other parameters of the system. The wall curvature balances the damping caused by the Hubble expansion and ensures that the network will reach a constant relativistic velocity.

The effect of the model parameters on the scaling solution is easy to infer. A larger $\sigma_3$ will make these walls more difficult to produce. An increase in the curvature $k$ will cause the resulting network to have a higher velocity $v_0$, and will also make the resulting densities smaller, since the increase in velocity will increase the probability that loops will form, dissipating energy. A faster rate of expansion will damp the system more, decreasing the velocity, and having the inverse effect on its densities.

Regarding energy losses, the value of $c$ determines the effectiveness and likelihood of loops being produced, therefore its increase will decrease the overall density, and will even out the amount of the three types of walls. Finally, the effectiveness of the interaction between walls of types 1 and 2 to form type 3 walls, described by $d$, affects the density of type 3 walls in relation to the other two types.

The other type of solution, with $k=0$, represents instead a system in which the domain walls are conformally stretched,
\begin{equation}
\alpha=\beta=\gamma=\lambda\,,\quad \delta=-3\lambda\,;
\end{equation}
in this scenario, the velocity decays, so the defects are effectively frozen by the expansion.

At a phenomenological level, this second solution corresponds to the case where the defects are locally planar and curvature is effectively confined to the junctions. In this `frustrated' case the walls are (locally) straight, therefore there is no mechanism for acceleration through curvature, and the walls will be frozen by the cosmological expansion. Other than this, the effects of the different parameters will be similar to the previous case.

\section{Exploring parameter space}
\label{numerics}

\begin{figure*}
\begin{center}
\includegraphics[width=2.7in]{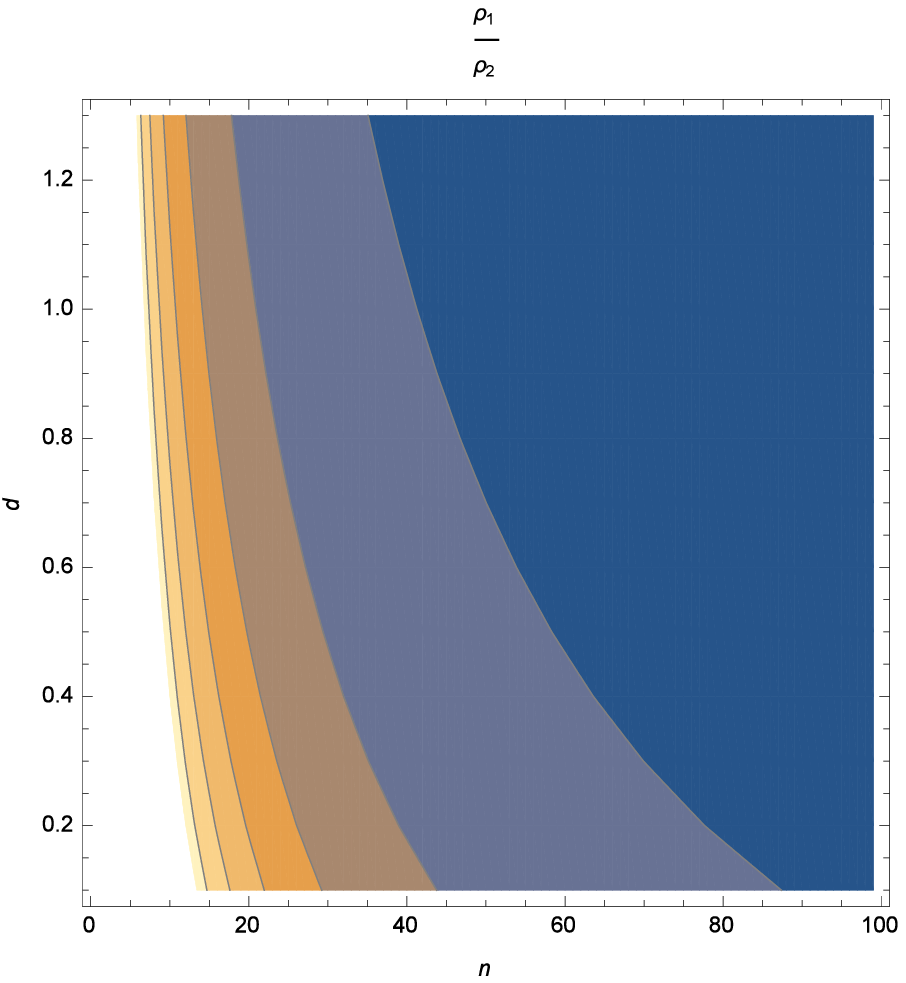}
\includegraphics[width=0.2in]{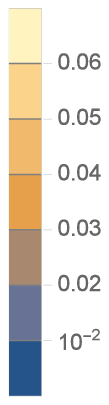}
\includegraphics[width=2.7in]{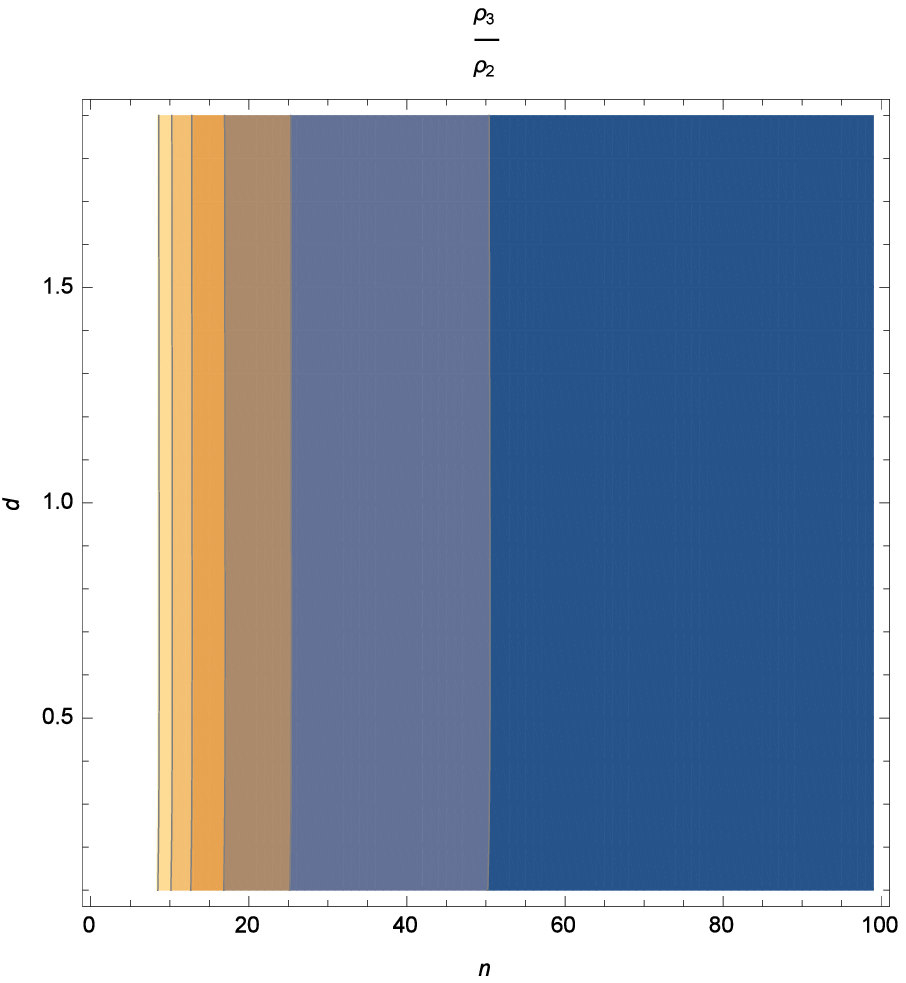}
\includegraphics[width=0.2in]{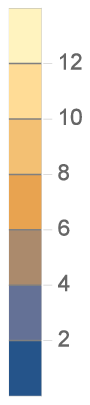}
\end{center}
\caption{\label{relativedensity}Contour plots of the density ratios of walls of types 1 and 2 (top panel) and 3 and 2 (bottom panel), as a function of $n=\sigma_2/\sigma_1$ (horizontal axis) and the interaction term $d$ (vertical axis).}
\end{figure*}

It is interesting to specifically discuss the behavior of the relative densities of the different types of walls. We will assume $\sigma_2=n\sigma_1$, fix $c$ and $k$ to their standard values, and plot the relative densities as a function of the free parameters $n$ and $d$. The plots presented in Figure \ref{relativedensity} are for $\lambda=0.7$, but one can verify that variations in the expansion parameter do not change the overall form of this dependence. The value of $\sigma_3$ was also fixed to $100\sigma_1$.

The results are consistent with intuitive expectations. In the first panel the walls with higher tension dominate the network, particularly for higher values of $d$ (when the rate of production of type-$3$ walls is higher). In the bottom panel we can see the tendency for type-$3$ walls to dominate the network when their tension is higher (though note that the network would take a very long time to reach this asymptotic state). For $n\approx 50$ there starts to be a balance between types 2 and 3, while the ones with lower tension (type 1) tend to disappear from the network.

\begin{figure*}
\begin{center}
\includegraphics[width=2.7in]{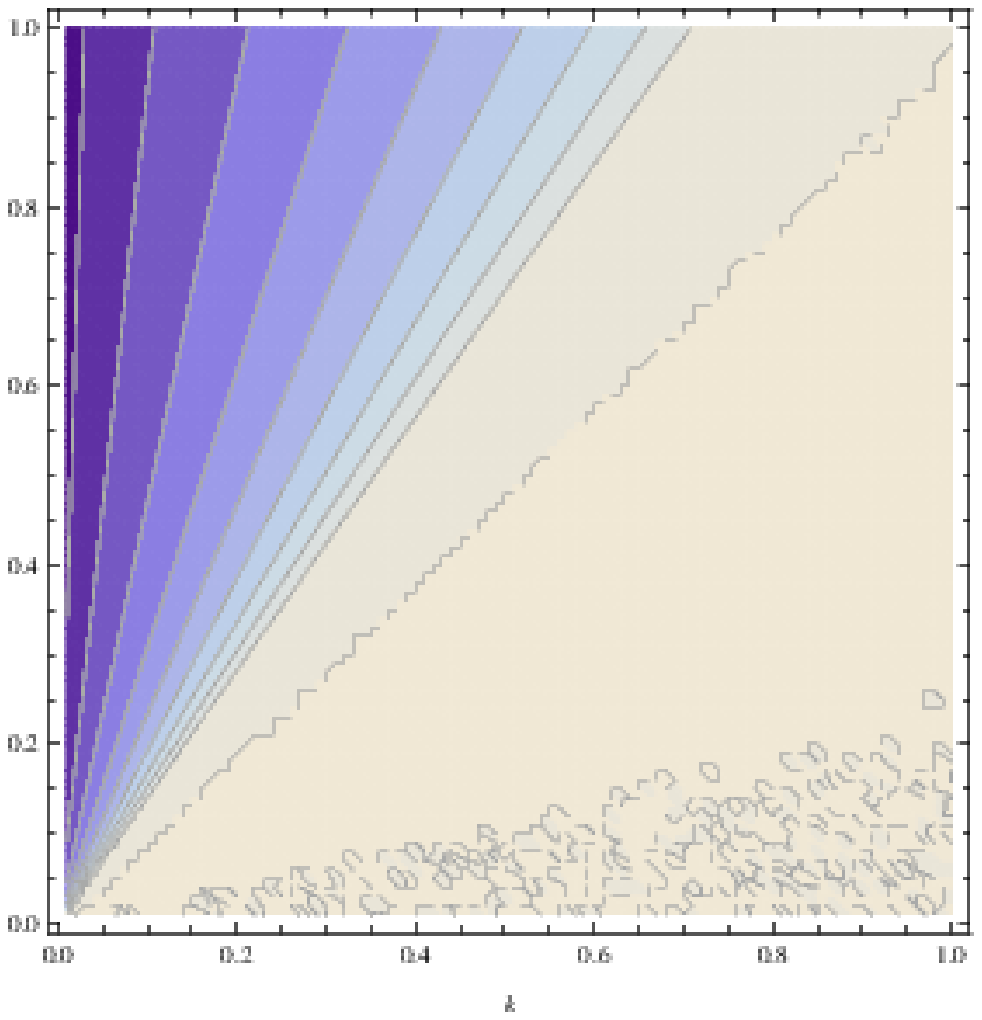}
\includegraphics[width=0.2in]{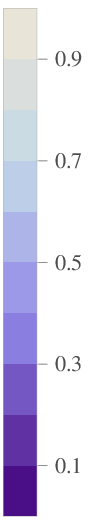}
\includegraphics[width=2.7in]{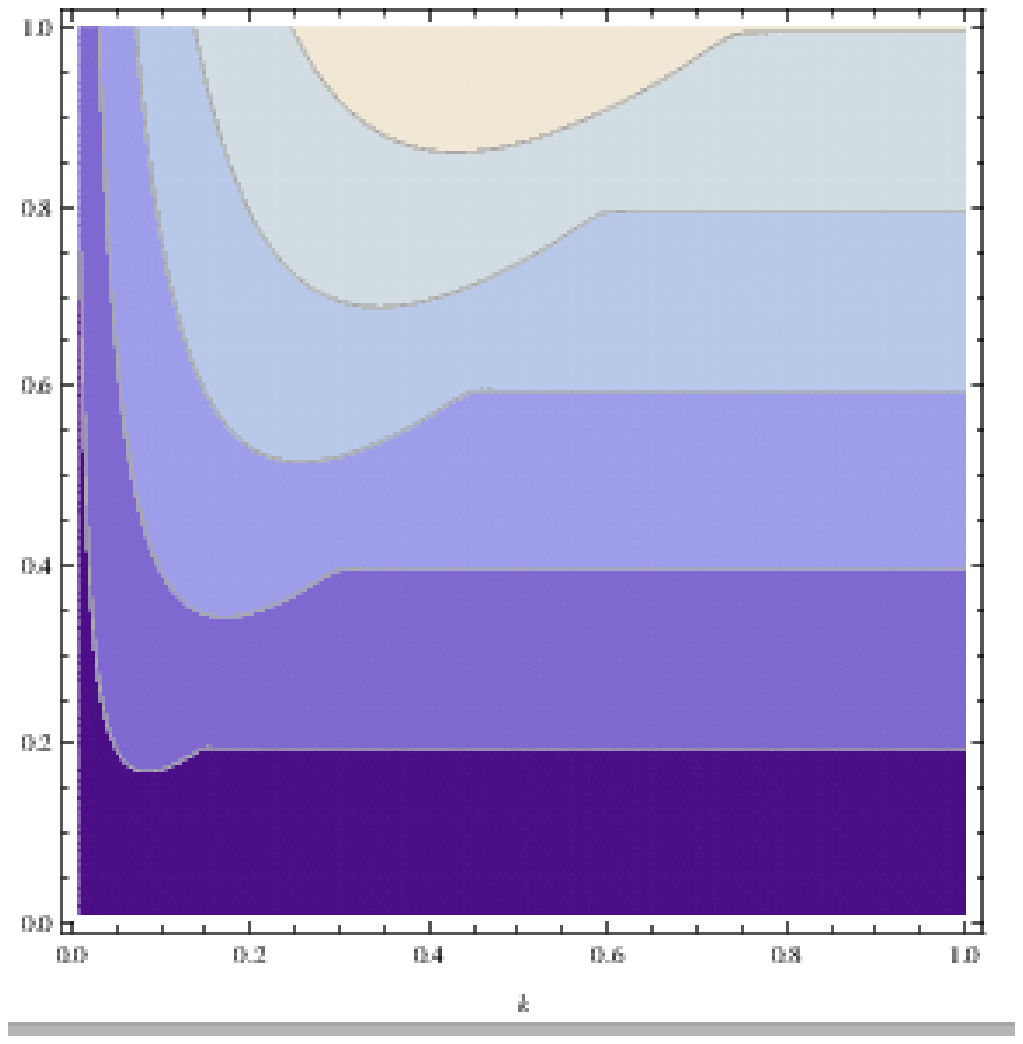}
\includegraphics[width=0.2in]{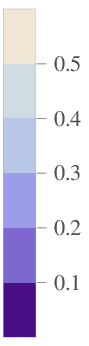}
\end{center}
\caption{\label{cont}Contour plots of the average velocity of the network (top panel) and of $\epsilon_1==L_1/t$ (bottom panel) as a function of the model parameters $k$ (horizontal axis) and $c$ (vertical axis).}
\end{figure*}

To further quantify how the choice of the parameters of the system influences the resulting scaling solution, we can obtain contour plots of the value of the functions $\epsilon_i=L_i/t$ and $v$, as function of the parameters $k$ and $c$, for ${c,k}\in]0,1]$. For simplicity we assume the limit $\sigma_1=\sigma_2=\sigma_3$ here. Figure \ref{cont} summarizes this result for the linear scaling solution branch; for these plots we have assumed a radiation-dominated era ($\lambda=0.5$) and an interaction parameter $d=0.1$.

This analysis confirms the results of the previous section and allows us to check the range of physically acceptable values for the different parameters. For example, rejecting solutions where the network reaches $v_0\to1$, the region in the bottom-right of the top panel of Figure \ref{cont} is excluded, narrowing the possible values for the respective parameters.

\begin{figure*}
\begin{center}
\includegraphics[width=2.7in]{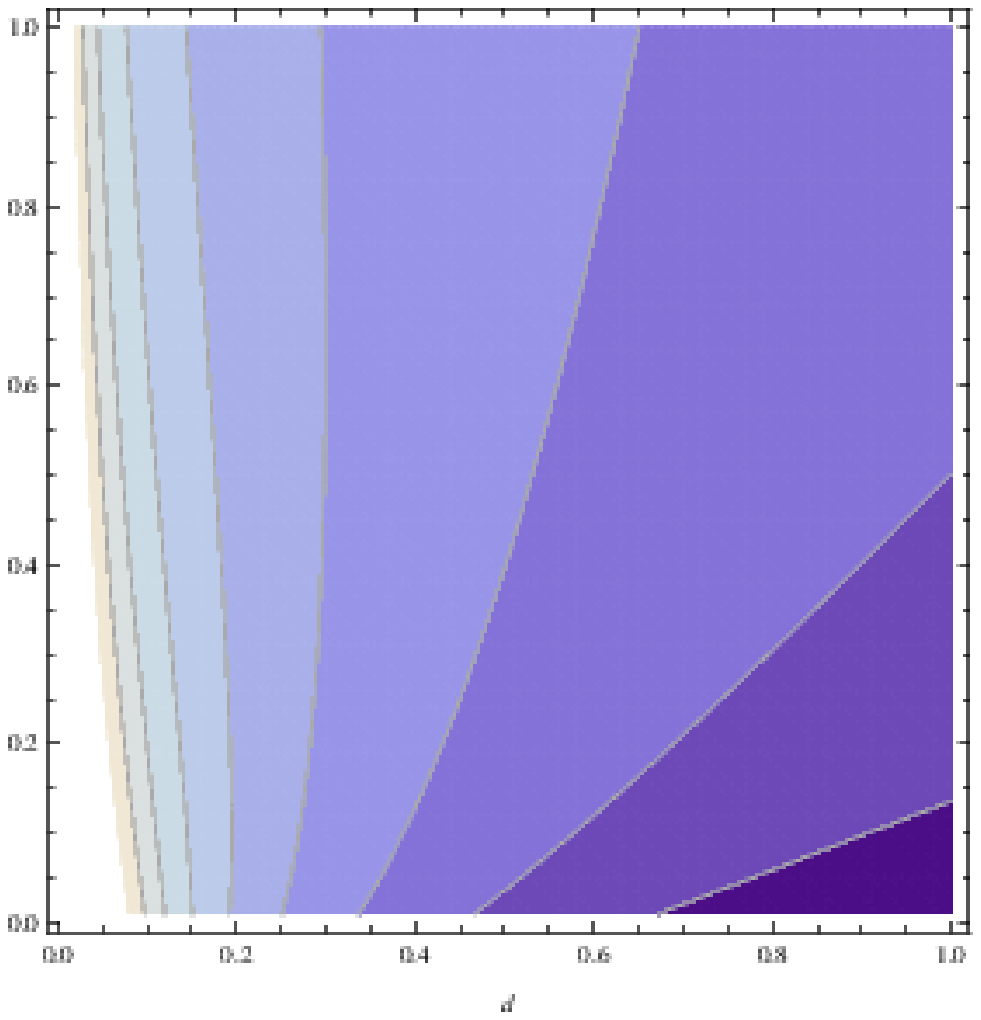}
\includegraphics[width=0.2in]{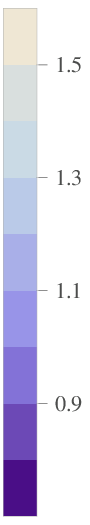}
\includegraphics[width=2.7in]{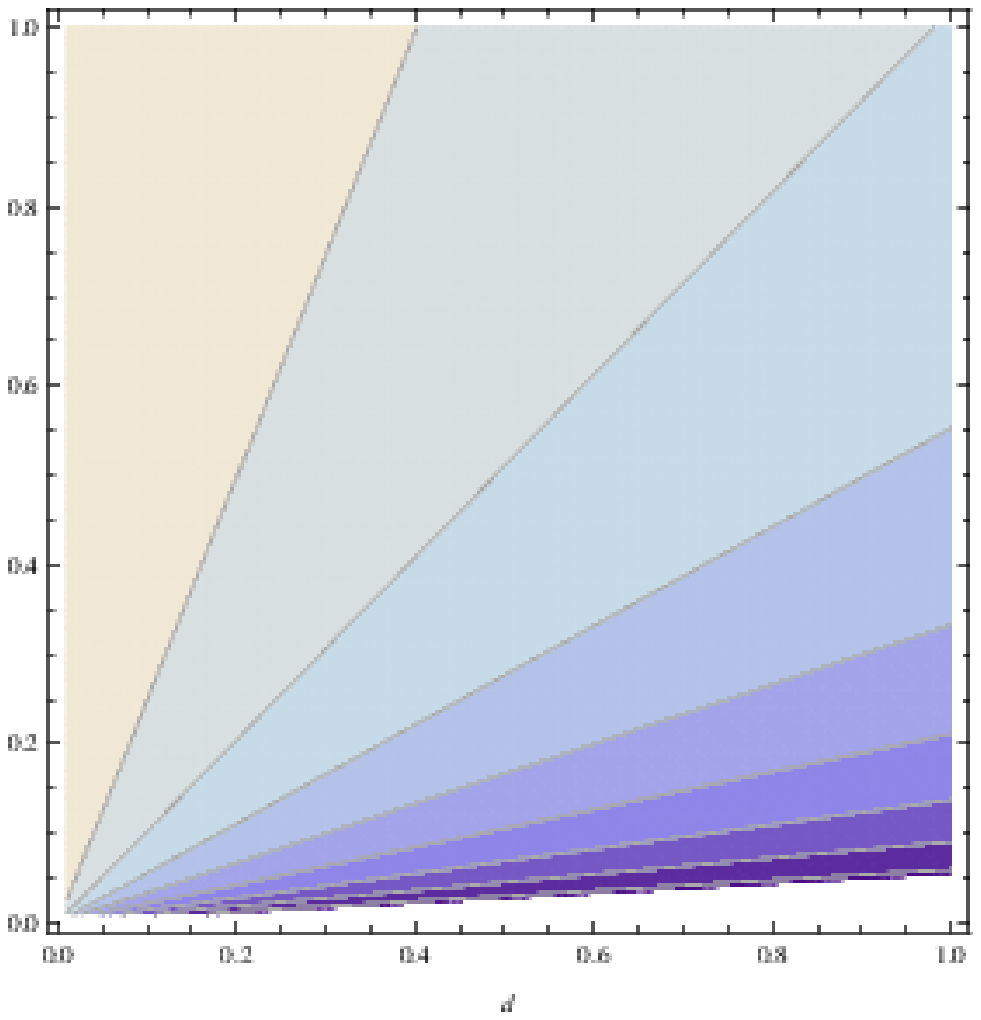}
\includegraphics[width=0.2in]{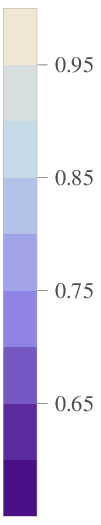}
\end{center}
\caption{\label{cont2}Contour plots of the ratio of the network correlation lengths $L_3/L_1$ as a function of $c$ (vertical axis) and $d$ (horizontal axis) for case $k=0$ (top panel) and $k=1/2$ (bottom panel).}
\end{figure*}

The $k\neq0$ and $k=0$ branches can be compared with the relevant free parameters being $c$ and $d$; this can be found in Figure \ref{cont2}. We can identify the behavior described previously, as with the increase in $d$ the quantity $L_3/L_1$ decreases for both cases, and $c$ has an effect that is much more noticeable in the case with wall curvature, since the loop term is proportional to the velocity.

\begin{figure*}
\begin{center}
\includegraphics[width=2.7in]{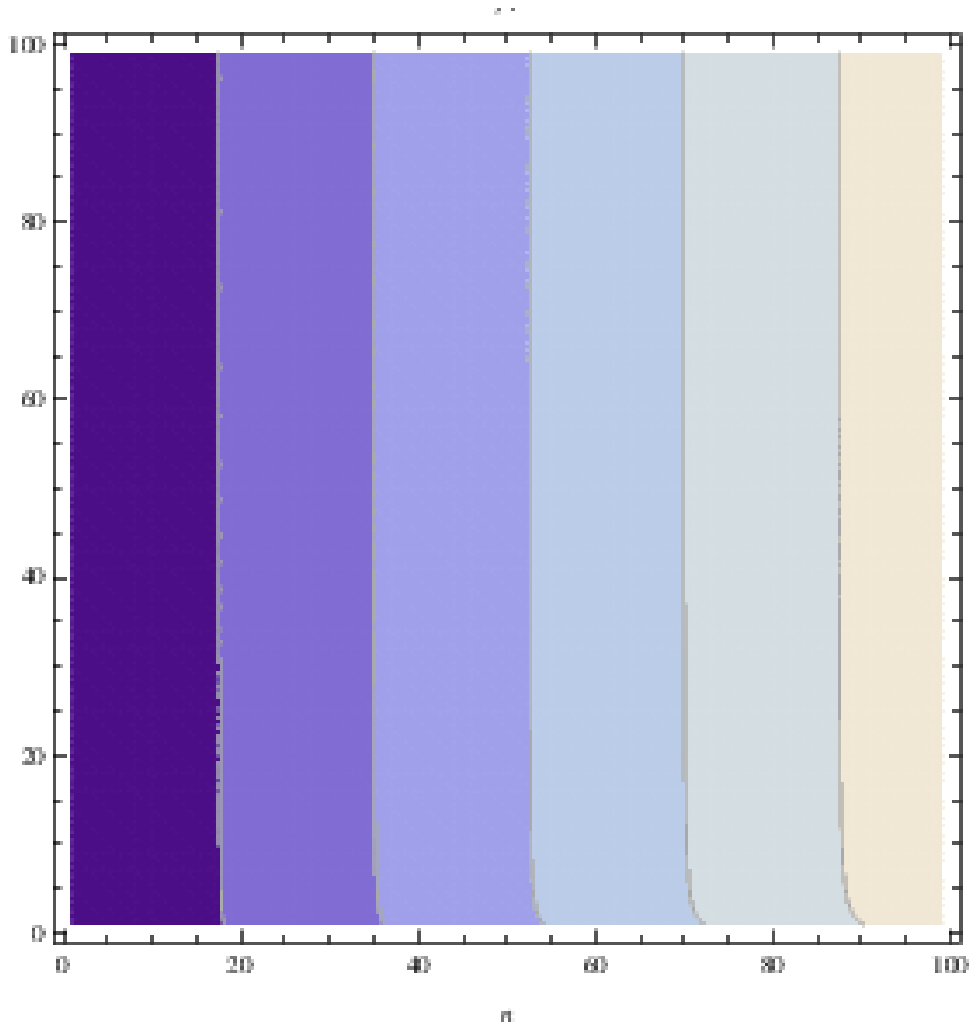}
\includegraphics[width=0.2in]{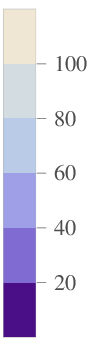}
\includegraphics[width=2.7in]{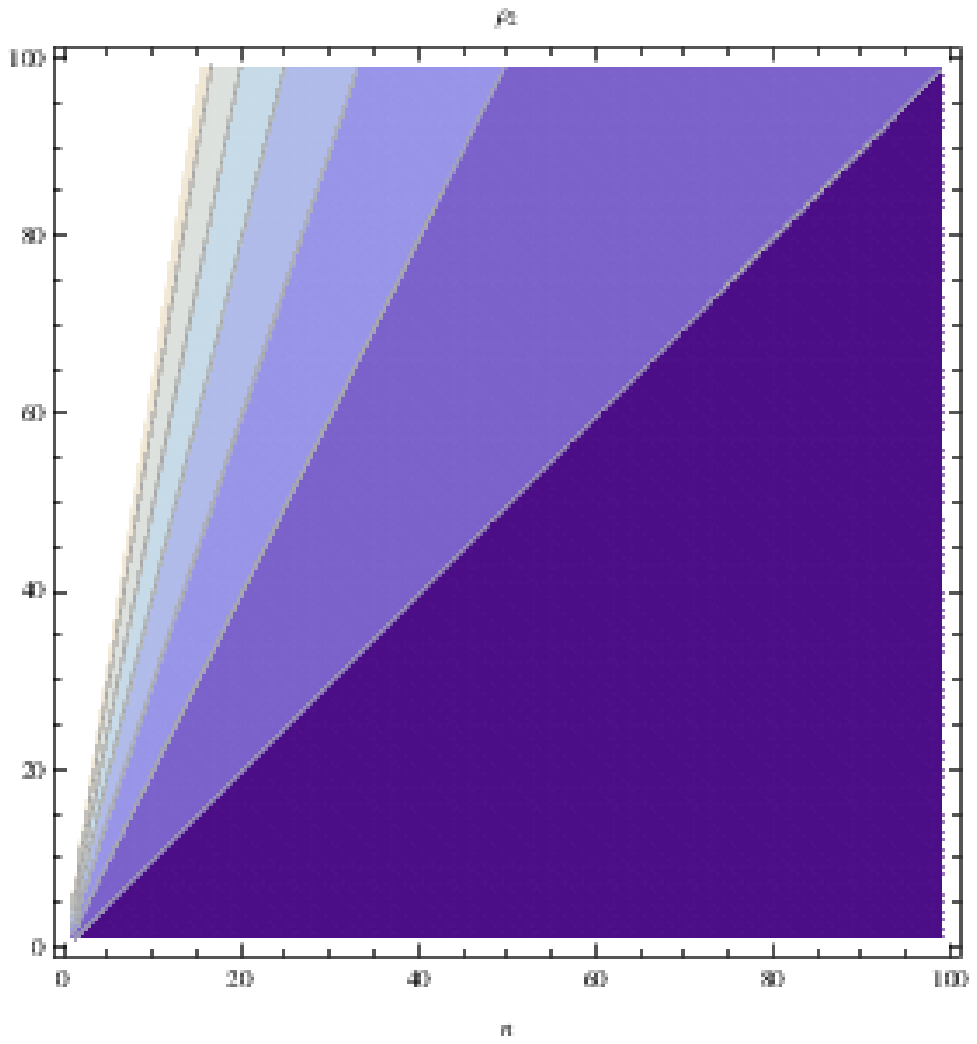}
\includegraphics[width=0.2in]{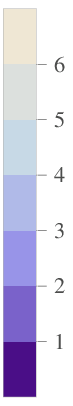}
\end{center}
\caption{\label{relativedensity2}Contour plots of the relative densities of walls $\rho_2/\rho_1$ (top panel) and $\rho_3/\rho_2$ (bottom panel) as a function of the parameters $m$ (vertical axis) and $n$ (horizontal axis).}
\end{figure*}

Finally we fix $c$ and $k$ to the values inferred from numerical simulations and generically define $\sigma_2=n\sigma_1$ and $\sigma_3=m\sigma_1$, and plot dimensionless density ratios for the network as a function of these new parameters, fixing some reasonable $d$. In Figure \ref{relativedensity2} we have done that, fixing $d=0.1$ and $\lambda=0.5$ (radiation epoch). The behavior is again consistent with intuitive predictions, since in the first panel the ratio is almost not affected by the parameter $m$ (except for high enough values of $d$, as expected). In the second panel, it is obvious the existence of a line $n=m$, which separates the regions of dominance of walls of types 2 and 3.

\section{Conclusions}
\label{conc}

We have explored a simple model for the evolution of a multi-tension domain wall network, containing three types of interacting walls. By studying the asymptotic scaling properties of these networks in generic expanding universes we have confirmed two classes of solutions: one corresponds to the well-known linear scaling solution, while the other corresponds to the case where the curvature is concentrated in the wall junctions. Our results complement those for cosmic strings \cite{Tye,Tasos}, though we emphasize that the way we model the network is slightly different and aims to model the properties that are more easily measurable in field theory numerical simulations.

Our results support the notion that linear scaling is a generic attractor solution for defect networks in a broad range of scenarios. On the other hand, we have also confirmed that concentrating the defect curvature on the defect junctions would be a possible way to obtain a conformally stretched network. While such 'frustration' scenarios \cite{Bucher,Frustrated} are quite tightly constrained \cite{Junctions}, this highlights a possible physical mechanism that would in principle lead to it. Further exploring this scenario will require high-resolution numerical simulations, which we leave for subsequent work.

%%%%%%%%%%%%%%%%%%%%%%%%%%%%%%%%%%%%%%%%%%%%%%%%%%%%%%%%%%%%%%%%%%%%%%%%%%%%%%%%%%%%%%%%%%%%%%%%%%%%%
\section*{Acknowledgements}
We acknowledge the financial support of grant PTDC/FIS/111725/2009 from FCT (Portugal). CJM is also supported by an FCT Research Professorship, contract reference IF/00064/2012, funded by FCT/MCTES (Portugal) and POPH/FSE (EC).

\end{document}